\documentclass[12pt]{article}
\usepackage{amsfonts,amssymb}

\date{March 11, 2001}

\def\be{\begin{equation}}
\def\ee{\end{equation}}
\def\bear{\begin{eqnarray}}
\def\eear{\end{eqnarray}}
\def\nn{\nonumber}

\def\wdg{{\wedge}}                              

\def\Im{{\rm Im\hskip0.1em}}

\newcommand{\px}[1]{{\partial_{#1}}}

\newcommand{\rep}[1]{{{\bf {#1}}}}      
\newcommand{\tr}[1]{{\mbox{tr}\{{#1}\}}}          
\newcommand{\ev}[1]{{\langle {#1} \rangle}}             

\def\BZ{\mathbb{Z}}

\newcommand{\MR}[1]{{\mathbb{R}^{#1}}}               
\newcommand{\MC}[1]{{\mathbb{C}^{#1}}}               
\newcommand{\MS}[1]{{{\bf S}^{#1}}}               
\newcommand{\CP}[1]{{\mathbb{C}{\mathbf{P}^{#1}}}}              
\newcommand{\AdS}[1]{{{\bf AdS}_{#1}}}            

\newcommand{\SUSY}[1]{{{\cal N}= {#1}}}           
\newcommand{\Spin}[1]{{{Spin}({#1})}}      

\def\a{{\alpha}}
\def\b{{\beta}}
\def\g{{\gamma}}

\def\u{{\mu}}
\def\v{{\nu}}
\def\th{{\theta}}

\def\hM{{M}}

\def\hU{{\hat{U}}}
\def\hO{{\hat{O}}}

\def\wstar{{\widetilde{\star}}}
\def\bk{{\overline{k}}}

\def\mcr{\mathcal{R}}
\def\xv{\vec{x}}
\def\xvt{\vec{x}^{\top}}
\def\nv{\hat{n}}
\def\nvt{\hat{n}^{\top}}
\def\Ot{\Omega^{\top}}

\def\cO{{\mathcal{O}}}

\def\mco{\mathcal{O}}
\def\kv{{\vec{k}}}

\def\dipL{\tilde{L}}   
\def\dmx{{\hat{x}}}    
\def\gYM{g_{\mbox{\tiny YM}}} 
\def\cpl{\lambda}  
\def\sbd{{\cal S}} 
\def\ngb{{\cal U}} 

\begin{document}

\begin{titlepage}
\titlepage
\rightline{hep-th/0103090, PUPT-1979}
\rightline{\today}
\vskip 1cm
\centerline{{\Huge Nonlocal Field Theories and their Gravity Duals}}
\vskip 1cm
\centerline{Aaron Bergman${}^{\clubsuit}$,
Keshav Dasgupta${}^{\diamondsuit}$,
Ori J. Ganor${}^{\clubsuit}$,
Joanna L. Karczmarek${}^{\clubsuit}$ and
Govindan Rajesh${}^{\diamondsuit}$}

\vskip 0.5cm

\begin{center}
\em ${}^{\clubsuit}$Department of Physics, Jadwin Hall \\
Princeton University \\
NJ 08544, USA
\end{center}

\vskip 0.5cm

\begin{center}
\em ${}^{\diamondsuit}$School of Natural Sciences \\
Institute of Advanced Study \\
Princeton, NJ 08540, USA
\end{center}

\vskip 0.5cm

\begin{center}
\tt
abergman,joannak@princeton.edu\\
origa@viper.princeton.edu\\
keshav,rajesh@sns.ias.edu
\end{center}

\abstract{
The gravity duals of nonlocal field theories in the large $N$ limit
exhibit a novel behavior near the boundary.
To explore this, we present and study the duals of dipole theories --
a particular class of nonlocal theories with fundamental dipole fields.
The nonlocal interactions are manifest in the metric of the gravity dual
and type-0 string theories make a surprising appearance.
We compare the situation to that in noncommutative SYM.
}
\end{titlepage}

\section{Introduction}\label{intro}
At the boundary,
the metric of $\AdS{d+1}$,
$ds^2 = \frac{\alpha'}{u^2}(dt^{2} - dx_1^2 -\cdots -dx_{d-1}^2- du^2)$,
becomes infinite.
This way the boundary is described by classical geometry, and
quantum gravity on $\AdS{d+1}$ can correspond to a local
field theory on a classical space \cite{Malda,GKP,WitAdS,AdSRev}.

The AdS/CFT correspondence can be extended to field theories on
a noncommutative space \cite{HasItz,MalRus,AOS}.
The gravity dual for the large $N$ limit of $\SUSY{4}$ Super-Yang-Mills
theory on a noncommutative $\MR{4}$ (NCSYM) with the noncommutativity in the
$2,3$ directions has the metric
\cite{HasItz,MalRus}
$$
ds^2 =\frac{\alpha'}{u^2}(dt^{2} - dx_1^2 - h(dx_2^2 +dx_3^2) - du^2)
$$
where $h= {u^4\over u^4+\theta^2}$ with $\theta = \theta_{23}$ the
typical length scale in the theory.
Here  the boundary, $u\rightarrow 0$, is no longer
classical. Indeed some components of the metric
tend to zero on the boundary.

Our motivation for this paper is to understand how nonlocality
in the field theory affects the metric of its gravity dual near the boundary.
Unfortunately, field theories on noncommutative spaces can be quite
complicated;
they exhibit UV/IR mixing and nonlocal behavior on varying scales.
UV/IR mixing,
    which means that high momentum is associated
with large-scale nonlocality and arbitrarily small momentum introduces
a new short-distance scale, can even obstruct the renormalization
procedure \cite{MSV,SV}.
Although $\SUSY{4}$ NCSYM is a finite theory and renormalizability is not
an issue, noncommutative geometry doesn't appear to be the simplest
way to introduce
nonlocality. There is a simpler way.

We will study a class of nonlocal gauge field theories
in which some of the fields correspond to dipoles of a constant
length.
Such theories were discussed in \cite{BerGan} in the context
of T-duality in noncommutative geometry. They were realized
in string theory in a different setting in \cite{DGG}.
Also see \cite{WitNGT,CGK,CGKM} for previous appearances of
such a theory.

At low energies these  ``dipole theories'' can be described as a
deformation of $\SUSY{4}$ SYM by a vector operator of
conformal dimension $5$.
This can be compared to the deformation by a tensor operator of conformal
dimension $6$ that describes NCSYM at low energy \cite{SWNCG,KZ,GRS}.
If the conformal dimension and the size of the Lorentz representation
is an indication of simplicity, then it is reasonable to expect that
dipole theories might be simpler than NCSYM.

The more interesting questions, however, hover in the UV region of the theory.
At distances shorter than the scale of the nonlocality,
we expect to find new phenomena.

Our ultimate goal is to answer the following questions:
\begin{itemize}
\item
How is the nonlocality of the dipole theory manifested in the
boundary metric?

\item
How does this manifestation of nonlocality compare to that of
noncommutative geometry? Are these features generic to the gravity
duals of nonlocal field theories?

\end{itemize}
In section ({\ref{nonloc}) we answer the first question for a
particularly simple dipole theory.
In the discussion, we show that this effect is analogous to a
feature of the supergravity dual of noncommutative geometry. We also
make some comments about the nature of the supergravity dual
for generic nonlocal theories.

The particular dipole theory that we study breaks supersymmetry
entirely. We chose to work with it because the supergravity equations
are simplified. The fermionic degrees of freedom, however, require
extra care. As we will argue, type-0 string theory with a strong
RR field strength has to be used
in order to correctly describe the gravity dual.

The paper is organized as follows.
In section (\ref{DipTh}) we review the construction of
dipole theories. In section (\ref{strth}) we describe
a simple string theory realization of these theories and
then calculate their gravity dual in section (\ref{sugsol}).
In section (\ref{ftcomp}) we compare the gravity dual and the
field theory in the infrared.
In section (\ref{geom}) we study the geometry of the gravity dual.
In section (\ref{nonloc}) we demonstrate the nonlocality
of the boundary. In section (\ref{fermions}) we discuss a puzzle
related to the nonlocal behavior of the fermions and argue
that type-0 string theory has to be invoked to resolve it.
In section (\ref{corrfcn}) we compute some
correlation functions and show how they exhibit some generic features
of nonlocality. Finally, in section (\ref{disc}) we discuss how the
features we have found here might
be generic to the supergravity duals of all nonlocal field theories.


\section{Dipole Theories}\label{DipTh}
Dipole theories are nonlocal field theories that also break Lorentz
invariance. They were obtained in \cite{BerGan} by studying the
T-duals of twisted fields
in noncommutative gauge theory. Below, we will describe how
to make a dipole theory out of an ordinary field theory.

\subsection{Definition}\label{dipdef}
We start with a local and Lorentz invariant field theory in $d$
dimensions. In order to turn it into a nonlocal theory we assign
to every field $\Phi_a$ a vector $L^\u_a$ ($\u=1\dots d$).
We will call this the ``dipole vector'' of the field.

The fields $\Phi_a$ can be scalars, fermions, or have higher spin.
Next, we define a noncommutative product
\be\label{defwst}
(\Phi_1\wstar\Phi_2)_{x} \equiv
\Phi_1(x -\frac{1}{2}L_2)\Phi_2(x+\frac{1}{2}L_1).
\ee
It is easy to check that this defines an associative product provided
that the vector assignment is additive, that is, $\Phi_1\wstar\Phi_2$
is assigned the dipole vector $L_1+L_2$.
For CPT symmetry, we will require that if $\Phi$ has
dipole vector $L$ then the charge conjugate field, $\Phi^\dagger$,
is assigned the dipole vector $-L$.
We will also require that gauge fields have zero dipole length.

In order to construct the Lagrangian of the dipole theory we need to replace
the ordinary product of fields with the noncommutative $\wstar$-product
(\ref{defwst}). In general, there might be some ordering ambiguity,
but the theories we will consider below are $SU(N)$ gauge theories
and have a natural ordering induced from the noncommutative products
of $N\times N$ matrices.

We have seen that the requirement of associativity translates into
a requirement of additivity for the dipole vectors.
One way to ensure this
is to have a global conserved charge in
the theory such that a field $\Phi_a$ has charge $Q_a$.
We then pick a constant vector $L^\u$ and assign to every field
$\Phi_a$ ($a=1\dots n$, where $n$ is the number of fields in the
theory) the dipole vector $Q_a L^\u$.
More generally, we can have $m$ global charges such that
a field $\Phi_a$ has the charges $Q_{j a}$ ($j=1\dots m$).
We can then pick a constant $d\times m$ matrix $\Theta^{\u j}$
($\u=1\dots d$ and $j=1\dots m$) and assign the field $\Phi_a$
a dipole vector $\sum_{j=1}^m \Theta^{\u j} Q_{j a}$.

Extending this definition by allowing $Q_a$ to be the momentum
we see that noncommutative Yang-Mills theory can also be
thought of as a dipole theory. The matrix $\Theta^{\u j}$
then becomes $\Theta^{\u\v}$ ($\v=1\dots d$) and is required
to be antisymmetric.
The dipole lengths are then both proportional to and
transverse to the momentum \cite{BigSus, Yin, ShJ}.

\subsection{A Dipole Deformation of $\SUSY{4}$ SYM}\label{symdef}

The dipole theories that we study in the rest of this paper
can be obtained from ordinary $SU(N)$ $\SUSY{4}$ SYM in 3+1D by turning
the scalars and fermions into dipole fields.
$\SUSY{4}$ SYM has 6 real scalars in the representation $\rep{6}$
of the R-symmetry group $SU(4)$ and 4 Weyl fermions in the representation
$\rep{4}$ of $SU(4)$.
We will use the global R-symmetry charges  to determine the
dipole vectors of the various fields as follows.
Pick $3$ constant commuting elements $V^\u\in su(4)$
($\u=1\dots 3$ and we will not consider time-like dipole vectors in
this paper), where $su(4)$ is
the Lie algebra of $SU(4)$. 
Take $V^\u$ to have dimensions of length.
Denote the matrix elements of $V^\u$
in the representation $\rep{4}$ as $\hU^\u_{j\bk}$ ($j,\bk =1\dots 4$).
Here $\hU^\u$ is a traceless Hermitian $4\times 4$ matrix.
Denote the matrix elements of $V^\u$ in the representation
$\rep{6}$ as $\hM^\u_{ab}$ ($a,b=1\dots 6$).
$\hM^\u$ is a real antisymmetric $6\times 6$ matrix.

Let $u^{(l)}_a$ ($a,l=1\dots 6$)
be an eigenvector of $\hM^\u$ with (real) eigenvalue $\dipL^\u_l$
so that $\sum_b\hM^\u_{ab} u_b^{(l)} = \dipL^\u_l u_a^{(l)}$.
$u^{(l)}_a$ does not depend on $\mu$ because
$[\hM^\u,\hM^\v] = 0$.
Let $\phi_a$ ($a=1\dots 6$) be the 6 real scalar fields of
$\SUSY{4}$ SYM.
Then the complex valued scalar fields
$\phi^{(l)}\equiv \sum_a u_a^{(l)}\phi_a$ are assigned a dipole vector
with components $2\pi \dipL^\u_l$ ($\u=1\dots d$).
Similarly, the fermionic fields are assigned dipole vectors that
are determined by the eigenvalues of the matrices $\hU^\u$.

\subsection{Supersymmetry}\label{DipSus}
The dipole theories obtained from $\SUSY{4}$ SYM in the previous
subsection are parameterized by $d$ constant traceless
Hermitian $4\times 4$ matrices $\hU^\u$.
For simplicity we will set $\hU^1=\hU^2=0$
and $\hU\equiv\hU^3$. Thus, the dipole vectors are all
in the $3^{rd}$ direction.
The matrix $\hU$
has dimensions of length, and its eigenvalues determine
the dipole vectors of the various fields.
Let the eigenvalues be $\a_1,\a_2,\a_3,-(\a_1+\a_2+\a_3)$.
Then, the dipole vectors of the various scalar fields are given by
$\pm (\a_i+\a_j)$ ($1\le i<j\le 3$).

The number of supersymmetries that are preserved by the dipole theory
is determined by the rank $r$ of $\hU$:
\begin{itemize}
\item
If $r=4$, then the theory is not supersymmetric at all.
\item
If $r=3$, there is one zero eigenvalue that we take by convention
to be $\a_3=0$, and the theory has $\SUSY{1}$ supersymmetry.
\item
If $r=2$, there are two zero eigenvalues that we take to be
$\a_2=\a_3=0$. The theory then has $\SUSY{2}$ supersymmetry.
The vector multiplet of $\SUSY{4}$ SYM decomposes as a vector
multiplet and a hypermultiplet of $\SUSY{2}$ SYM.
All the fields in the $\SUSY{2}$ vector multiplet have dipole vector
$0$, and the fields in the hypermultiplet have dipole vectors $\pm\a_1$.
\end{itemize}

Because we can realize dipole theories without supersymmetry, one
might ask if poles similar to those discovered in \cite{MSV,SV} might arise in
the perturbative expansion of the theory. In fact, they do not. This
can be seen by examining the expression of \cite{MSV,SV} 
for the effective cutoff
$$
\Lambda_{\mathrm{eff}} \rightarrow
\frac{1}{\sqrt{\Lambda^{-2} + (\theta p)^{2}}}.
$$
We recognize $\th p$ as the length of the dipoles in noncommutative
geometry. Thus, the analogous expression in our theory is
$$
\Lambda_{\mathrm{eff}} \rightarrow
\frac{1}{\sqrt{\Lambda^{-2} + L^{2}}}
$$
which, as it is independent of the momenta, gives rise to no new poles.

\section{String Theory Realization of Dipole Theories}\label{strth}
In order to find the gravity dual of the large $N$ limit of a
particular dipole theory, we need to find a simple string theory realization
for it. We now do this for a large class of dipole theories.

In \cite{DGG}, a realization of dipole theories
with $\SUSY{2}$ supersymmetry was suggested using D3-branes that
probe the center of a modified Taub-NUT geometry.
While this realization is convenient for a BPS analysis it is hard
to extract the gravity dual from it, and it is not obvious how
to generalize it to dipole theories that break $\SUSY{2}$ supersymmetry.

Fortunately, the Taub-NUT space that was used in \cite{DGG} is not
an essential ingredient. We can find an alternative setting that
has the same behavior near the brane probes.
This setting, which we will describe below, has the disadvantage
that the geometry is not asymptotically Euclidean at infinity.
Nevertheless, it has been constructed in string theory 
\cite{RusTse} and is good for extracting the gravity duals that we seek.
Other worldsheet CFTs that break Lorentz invariance have been studied in
\cite{dBS}.


The backgrounds that we consider are twisted versions of type-II string
theory. They are related to the Melvin solution \cite{Melvin} and are
in fact identical to the backgrounds discussed in \cite{DGGH} and more recently
in \cite{CosGut,Str}.
As was shown in \cite{DGGH}, the twisted backgrounds are unstable,
and the instability is similar to that discussed in \cite{WitKK}.
This instability is exponentially suppressed as $g_s\rightarrow 0$ and
is likely to be completely absent when some supersymmetry is preserved.
For the time being we will ignore the instability. We will return to this
point in the discussion.

\subsection{The T-dual of a Twist}\label{TDtw}
We will first describe a type-II background without branes and
then later we will add the brane probes.
Consider type-IIA string theory on a space that is 
$\MR{9,1}$ modded out by the isometry
$$
{\cal U}: (x_0,x_1,x_2,x_3,\{x_{3+a}\}_{a=1}^6)
\mapsto
(x_0,x_1,x_2,x_3 + 2\pi R_3, \{\sum_{b=1}^6 O_{ba} x_{3+a}\}_{a=1}^6).
$$
Here $O\in SO(6)$ is an orthogonal matrix.
The twisted compactification is parameterized by $R_3$ and,
because we need to define the action on fermions, an element
of $\Spin{6} \cong SU(4)$.
This background is, in general, modified by quantum corrections,
but $O$ and $R_3$ are defined by their asymptotic values at infinity.
We will denote this background by $X(O,R_3)$.
Note that if $R_3 > 0$ the isometry ${\cal U}$
has no fixed points and therefore $O$ is not necessarily of finite order.

Now consider probing $X(O,R_3)$ with D2-branes in directions
$(x_0,x_{1},x_2)$ and then taking the limit $R_3\rightarrow 0$
together with
$O = e^{\frac{2\pi i R_3 \hM}{\a'}}$ where $\hM$ is a finite matrix of
the Lie algebra $so(6)\cong su(4)$ with dimensions of length and $\frac{\a'}{2\pi}$
is the inverse string tension.

When $\hM=0$, we can perform T-duality to
transform the D2-branes into D3-branes.
When $\hM\neq 0$, we will now show the low energy description
of the probe is a dipole theory.

\subsection{Branes Probing Dual Twists}\label{brprob}
We wish to find the low energy Lagrangian describing D2-branes that
probe the twisted geometry of subsection (\ref{TDtw}).
The light degrees of freedom come from the strings with two
Dirichlet boundary conditions, {\it i.e.\/}, fundamental
strings with ends on the D2-branes. Because $R_3\rightarrow 0$, we
have to set the string oscillators to their ground states, but
the winding number can be arbitrary.

To obtain the Lagrangian, we can adopt a procedure similar
to the one described in \cite{CheKro,SJi,ChuHo} for noncommutative gauge
theories.
Also, the construction that we present here is reminiscent of
the construction in \cite{DH}.
In momentum space, the action of the dipole theory is obtained
from the action of $\SUSY{4}$ SYM by inserting certain phases.
Let $\Phi_1(p_1),\dots,\Phi_n(p_n)$
be fields in the adjoint representation
of $U(N)$ and suppose that $\SUSY{4}$ SYM has a term of the form
\be\label{trphiphi}
\tr{\Phi_1(p_1)\cdots \Phi_n(p_n)},
\ee
in the Lagrangian (of course $n\le 4$).
The variables $p_i$ are the momenta.
Let the dipole vectors of the fields be $L_1,\dots, L_n$.
We have
$$
\sum_{i=1}^n L_i=0,\qquad
\sum_{i=1}^n p_i=0.
$$
The dipole theory is obtained from the ordinary $\SUSY{4}$ SYM theory
by inserting the phases
\be\label{addphase}
e^{i\sum_{1\le i<j\le n} p_i L_j}
\ee
in front of terms like (\ref{trphiphi}).
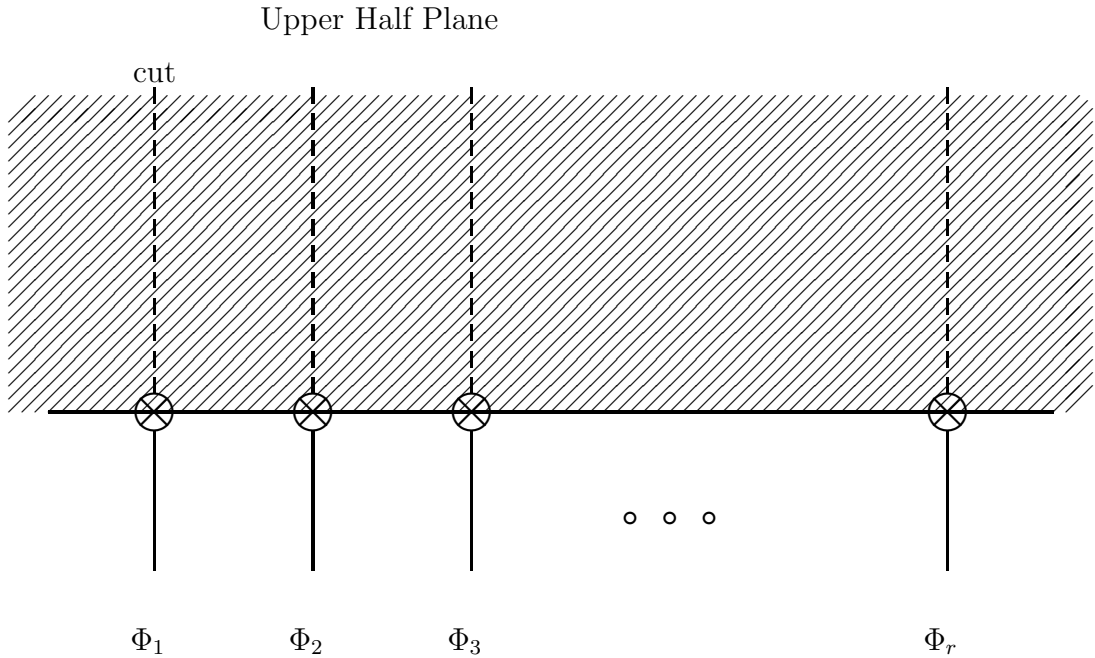
\begin{figure}[t]
\begin{picture}(400,255)
\thicklines
\put(20,100){\line(1,0){380}} 
\thinlines
\multiput(5,100)(5,0){58}{\line(1,1){120}} 
\put(295,100){\line(1,1){118}} 


\newcounter{Yco}
\newcounter{TLen}
\setcounter{Yco}{105}
\setcounter{TLen}{115}
\loop
\put(5,\value{Yco}){\line(1,1){\value{TLen}}}
\addtocounter{Yco}{5}
\addtocounter{TLen}{-5}
\ifnum \value{TLen}>0
\repeat

\newcounter{Xco}
\setcounter{Xco}{300}
\setcounter{TLen}{115}
\loop
\put(\value{Xco},100){\line(1,1){\value{TLen}}}
\addtocounter{Xco}{5}
\addtocounter{TLen}{-5}
\ifnum \value{TLen}>0
\repeat

\put(100,245){Upper Half Plane}

\thicklines
\multiput(60,100)(60,0){3}{\circle{14}} 
\multiput(55,95)(60,0){3}{\line(1,1){10}}
\multiput(55,105)(60,0){3}{\line(1,-1){10}}
\multiput(60,40)(60,0){3}{\line(0,1){53}} 

\put(51,10){$\Phi_1$}
\put(111,10){$\Phi_2$}
\put(171,10){$\Phi_3$}

\multiput(240,60)(15,0){3}{\circle{4}} 


\put(360,100){\circle{14}} 
\put(355,95){\line(1,1){10}}
\put(355,105){\line(1,-1){10}}
\put(360,40){\line(0,1){53}} 

\put(351,10){$\Phi_r$}

\multiput(60,107)(0,10){12}{\line(0,1){6}} 
\multiput(120,107)(0,10){12}{\line(0,1){6}} 
\multiput(180,107)(0,10){12}{\line(0,1){6}} 
\multiput(360,107)(0,10){12}{\line(0,1){6}} 

\put(52,225){cut}

\end{picture}
\caption[Vertex Operators for the Twisted Theory]{\label{wsfig}An
$r$-point amplitude with vertex operators that
    carry winding numbers. It requires $r$ cuts on the worldsheet.}
\end{figure}
Now let us consider branes probing  $X(O,R_3)$.
For simplicity, let us assume that the twist, $O$, acts only on
$Z\equiv X^8 + i X^9$ as $Z\rightarrow e^{i\a} Z$.
We will refer to the angular momentum corresponding
to rotation in the $Z$-plane as the {\it $Z$-charge\/}.

In the case that $\a=0$ we know that the theory on the D2-brane
probe is $\SUSY{4}$ SYM. The states with momentum along the $3^{rd}$
direction, in the SYM theory, correspond to winding states
along the $3^{rd}$ direction in the string theory setting.

Now let us turn on the twist, $\a$.
Consider a string disc amplitude that calculates the interaction
of $n$ open string states with winding numbers $w_1,\dots,w_n$
and with {\it $Z$-charges\/} $q_1,\dots,q_n$.
The worldsheet theory has a global $U(1)$ symmetry corresponding
to the $Z$ charge.
The string vertex operators that correspond to the external states
are charged under this $U(1)$ symmetry.
The disc worldsheet has cuts that emanate from the external
vertex operators on the boundary. Along the $j^{th}$ cut,
the worldsheet field $Z$ jumps by a phase $e^{i \a q_j}$.
We can redefine the field $Z$ to be continuous, but then
there will be additional phases coming from the
vertex operators on the boundary.
It is easy to see that this phase is
$$
e^{i\a \sum_{1\le i<j\le n} w_i q_j}.
$$
This is illustrated in figure \ref{wsfig}.
It agrees with (\ref{addphase}) because
$$
L_j = \a'\frac{\a}{R_3}q_j,\qquad
p_j = {\a'}^{-1} R_3 w_j.
$$

In 3+1D, the photon of the $U(1)\subset U(N)$ center
of the gauge group is likely to become massive via a dynamical
mechanism similar
to the one described in \cite{DouMoo} for quiver theories, so that
the gauge group is actually just $SU(N)$, but we will
ignore this for the time being.

\section{Supergravity Solution for a Twisted Brane}\label{sugsol}
We now turn to the task of describing the supergravity duals of these
dipole theories.
We will use the string theory realization of dipole theories as
described in the previous section.

We will find the exact {\it classical\/} supergravity solutions in
four steps:
\begin{enumerate}
\item
We start with the D3-brane solution of type-IIB classical
supergravity and compactify one of the directions parallel to the
D3-branes. We will call it the $3^{rd}$ direction.
\item
We perform T-duality on the $3^{rd}$ direction to obtain
a solution that describes D2-branes in type-IIA.
The solution, however, will be translationally invariant
along the $3^{rd}$ direction, and, as such,
it describes smeared rather than localized D2-branes.
\item
We now insert a transverse $SO(6)$ twist into the geometry by hand.
This is accomplished by simply changing the boundary conditions for
the 6 transverse coordinates
as we complete a circle around the $3^{rd}$ direction.
Locally, the metric is unchanged.
\item
Finally, we use T-duality to turn the smeared D2-branes back to
D3-branes.
\end{enumerate}

In this paper we will restrict ourselves to cases where all the
dipole vectors of the fields in the theory are oriented in the same direction.
This was  direction $3$ above. In appendix (\ref{appthree}) we present
the generalization for generic dipole vectors.
We now turn to the details.

\subsection{The Type-IIB D3-Brane}
First, the conventions. We work in the $(+,-,\dots,-)$ metric. Greek
indices are $\u,\v= 0\dots 2$. The time direction is $t = x_{0}$. The
direction that we T-dualize is the $3^{rd}$.
The remaining directions,
perpendicular to the brane, are labeled by
roman indices $a,b = 4\dots 9$. All
metrics will be in string frame.

We start with the metric for a D3-brane (note that all the $x$'s have
dimensions of length)
$$
        ds^{2}_{\rm str} = H^{-\frac{1}{2}}(dt^{2}
        - dx_{1}^{2}- dx_{2}^{2} - dx_{3}^{2}) -
        H^{\frac{1}{2}}(\delta_{ab}dx^{a}dx^{b})
$$\noindent
where
$$
      H = 1+\frac{\mcr^{4}}{r^{4}},\qquad
\mcr^4 = 4\pi g_s N{\a'}^{2},\qquad
r^2=\delta_{ab}x^{a}x^{b},
$$\noindent
and we have the following backgrounds for the RR 4-from
potential and the dilaton respectively
$$
        C^{(4)}_{0123} = -H^{-1},\qquad
        e^{2\phi} = e^{2\phi_{0}}.
$$

Next, we compactify along $x^{3}$ with radius $R_3 \equiv R$. The metric is
now
$$
        ds^{2} = H^{-\frac{1}{2}}(dt^{2} -
        \delta_{\u\v}dx^{\u}dx^{\v} - R^{2} d\dmx_{3}^{2}) -
        H^{\frac{1}{2}}(\delta_{ab}dx^{a}dx^{b})
$$
Note that $\dmx_3$ is now dimensionless and periodic
$\dmx_3\sim \dmx_3+2\pi$.

\subsection{A Smeared D2-Brane with a Twist}
We now T-dualize around $x_3$. Following \cite{BUSH,BUSC,BHO}, we have
\bear
       & C^{(3)}_{012} = \frac{8}{3} H^{-1} \qquad
      e^{2(\phi-\phi_{0})} = \frac{\a'}{R^2}H^{\frac{1}{2}} & \nn\\
       & ds^{2} = H^{-\frac{1}{2}}\left(dt^{2} - dx_{1}^{2} - dx_{2}^{2}\right)
      -H^{\frac{1}{2}} \left(
   \frac{{\a'}^{2}}{R^{2}}d\dmx_{3}^2 + dx_{4}^{2}
    + \cdots + dx_{9}^{2}\right) &\nn
\eear
This is a smeared D2-brane.
We can now add a twist to the transverse directions $x_{4},\cdots,x_{9}$
as we travel around the circle $x_{3}$. In particular, we take
an element of the Lie algebra $so(6)$, $\Omega_{ab}$, and change the
metric to
$$
       ds^{2} = H^{-\frac{1}{2}}(dt^{2} - dx_{1}^{2} - dx_{2}^{2})
   - H^{\frac{1}{2}}\left\{{\a'}^{2}R^{-2}d\dmx_{3}^{2} +
   \sum_a \left(dx_{a} - \sum_{b}\Omega_{ab} x_{b} d\dmx_{3}\right)^{2}\right\}
$$
We can expand this out, giving
\bear
       ds^{2} &=& H^{-\frac{1}{2}}(dt^{2} - dx_{1}^{2} - dx_{2}^{2}) \nn\\
       &&- H^{\frac{1}{2}}\left\{( {\a'}^{2}R^{-2} + \xvt \Omega^{\top}
       \Omega \xv)d\dmx_{3}^{2} + d\xvt d\xv
       -2d\xvt\Omega \xv\ d\dmx_{3}\right\}\nn
\eear\noindent
where $\xv$ is the vector formed by $x_{a}$ ($a =4\dots9$).

\subsection{And Back to the D3-Brane}
Once again, we apply the T-duality formulae (recall that $\dmx_3$ is
dimensionless, while all the other coordinates have dimension of length)
\bear
        ds^{2} &=& H^{-\frac{1}{2}}\left(dt^{2} - dx_{1}^{2} - dx_{2}^{2}
   - \frac{{\a'}^2}{{\a'}^2 R^{-2} + \xvt\Omega^{\top}\Omega\xv}
d\dmx_{3}^2\right) \nn\\
        && - H^{\frac{1}{2}} \left(d\xvt d\xv - \frac{(d\xvt\Omega\xv)^{2}}
        {{\a'}^2 R^{-2} + \xvt \Omega^{\top}\Omega \xv}\right)\nn
\eear

We also have
\bear
         C^{(4)}_{3012} &=&  H^{-1}  \nn\\
         \sum_{a} B_{\hat{3}a} dx^{a} &=& - \frac{d\xvt \Omega \xv}
        {{\a'}^2 R^{-2} + \xvt\Ot\Omega\xv} \nn\\
        e^{2(\phi-\phi_{0})} &=& \frac{1}{1 +
{\a'}^{-2} R^{2} \xvt \Ot\Omega \xv}\nn
\eear\noindent
which we will address later on.

Defining $\xv = r\nv$ so that $\|\nv\| = 1$,
our metric becomes
\bear
        ds^{2} &=& H^{-\frac{1}{2}}\left(dt^{2} - dx_{1}^{2} - dx_{2}^{2}
        - \frac{{\a'}^2}{{\a'}^2 R^{-2} +
r^{2} \nvt\Ot\Omega\nv}d\dmx_{3}^2\right) \nn\\
        && - H^{\frac{1}{2}} \left(dr^{2} + r^{2} d\nvt d\nv
-\frac{r^4}{{\a'}^2 R^{-2} + r^{2}\nvt \Omega^{\top}\Omega \nv}
   (\nvt \Ot d\nv)^{2}
\right)
\nn
\eear

\subsection{The Near-Horizon Limit}
In these coordinates, the horizon is at $r=0$, and so the near-horizon
limit is $r$ small.
We can therefore approximate
$$
      H^{\frac{1}{2}} = \sqrt{1+\frac{\mcr^{4}}{r^{4}}} \sim
      \left(\frac{\mcr}{r}\right)^{2}.
$$

Substituting this into the metric, we obtain
\bear
        ds^{2} &=& (r/\mcr)^{2}\left(dt^{2} - dx_{1}^{2} - dx_{2}^{2}
    -\frac{{\a'}^2}{{\a'}^2 R^{-2} + r^{2} \nvt\Ot\Omega\nv} d\dmx_{3}^{2}
\right) \nn\\
        && -  \frac{\mcr^{2}}{r^{2}}dr^2 - \mcr^{2}\left(d\nvt d\nv -
        \frac{1}{{\a'}^2 (rR)^{-2} + \nvt \Omega^{\top}\Omega \nv}
(\nvt \Ot d\nv)^{2}
\right)\nn
\eear

Finally, we make the substitution
$u = \mcr^2/r$
\bear
ds^{2} &=& \frac{\mcr^2}{u^2}\left(dt^{2} - dx_{1}^{2} - dx_{2}^{2}
      - \frac{u^{2}}{(\frac{u}{R})^{2} + \cpl^2\nvt\Ot\Omega\nv} d\dmx_3^2\right)
      - \mcr^{2}\frac{du^{2}}{u^2}
\nn\\ &&
- \mcr^{2}\left(d\nvt d\nv -
        \frac{\cpl^2}{(\frac{u}{R})^{2} + \cpl^2\nvt \Omega^{\top}\Omega \nv}
      (\nvt \Ot d\nv)^{2}
\right)
\nn
\eear
where $\cpl^2\equiv \frac{\mcr^4}{{\a'}^2} = 4\pi \gYM^2 N$.
The dilaton and the NSNS 2-form field are
\bear
\sum_{a} B_{\hat{3} a} d\nv^a &=&
         - \frac{\cpl^2}{\frac{u^2}{R^2} + \cpl^2\nvt\Ot\Omega\nv} d\nvt \Omega \nv,
\nn\\
        e^{2(\phi-\phi_{0})} &=& 
  \frac{1}{1 + \frac{R^{2}}{u^2} \cpl^2 \nvt \Ot\Omega \nv}
\nn
\eear
Now we take the limit $R\rightarrow \infty$ keeping $R\Omega = \hM$ fixed.
We also redefine $\dmx_3 = \frac{x_3}{R}$.
Note that $x_3$, $u$ and $\hM$ have dimensions of length. We find
\bear
        ds^{2} &=& \frac{\mcr^2}{u^2}
     \left(dt^{2} - dx_{1}^{2} - dx_{2}^{2} -du^2
        - \frac{u^{2}}{u^{2} + \cpl^2\nvt\hM^\top\hM\nv} dx_3^2\right)
\nn\\ &&
- \mcr^{2}\left(d\nvt d\nv -
        \frac{\cpl^2}{u^2 + \cpl^2\nvt \hM^\top\hM \nv} (\nvt \hM^\top d\nv)^{2}
\right)
\label{twmet}
\eear
The NSNS 2-form field and the dilaton are
\bear
\sum_a B_{3 a} d\nv^a &=&
         - \frac{\cpl^2}{u^2 + \cpl^2\nvt\hM^\top\hM\nv}
d\nvt \hM \nv,
\nn\\
   e^{2(\phi-\phi_{0})} &=& \frac{u^{2}}{u^{2} + \cpl^2\nvt \hM^\top\hM \nv}
\label{Btw}
\eear

Given this form of the metric it is not obvious that the region
$\frac{\mcr}{r}\gg 1$ indeed decouples from the bulk, as we have assumed.
In principle, one can calculate scattering amplitudes for gravitons 
as in \cite{GKT,GubKle}. In some cases one can see from the scattering amplitudes
that the bulk does not decouple (see for instance \cite{AIO}).

In our case, the geometry is strongly coupled when $u$ is small, as will be
discussed in more detail in subsection (\ref{xtdl}), and evaluating the
scattering amplitudes is difficult.
Nevertheless, there is no reason to expect that the bulk will not decouple.
The dipole-theories describe well defined renormalizable theories that
do not require additional degrees of freedom in the UV.

\section{Comparison to the Field Theory in the Infrared}\label{ftcomp}

For $\hM = 0$, the metric (\ref{twmet}) describes
$\AdS{5}\times \MS{5}$ with boundary at $u = 0$.
The IR region corresponds to large $u$.
For large $u$ the deviation from the standard $\AdS{5}\times\MS{5}$
metric
describes a deformation of $\SUSY{4}$ SYM by irrelevant operators.

In \cite{BerGan}, the first order
correction to the SYM Lagrangian was determined as the dipole length times
the dimension 5 operator
\bear
{\cal O}_\u^{IJ} &=&
       {i\over {\gYM^2}}
       \tr{{F_{\u}}^\v\Phi^{[I} D_\v\Phi^{J]}
   +\sum_{K}(D_\u\Phi^K)\Phi^{[K}\Phi^I\Phi^{J]}}
        + {\mbox{fermions}} \nn
\eear
Here $I,J=1\dots 6$ are R-symmetry indices,
$\Phi^I$ ($I=1\dots 6$) are the scalars,
$D_\u = \px{\u} -i [A_\u,\cdot]$ is the
covariant derivative, $F_{\u\v}$ is the field strength
and $[\cdots ]$ means complete anti-symmetrization.
${\cal O}_\u^{IJ}$
is a vector operator that transforms in the $\rep{15}$ of the
R-symmetry group $SU(4)$.

We should be able to find the dual of this in the supergravity.
The only vector field on $\AdS{}$ that we have obtained is the two-form
$$
\sum_a B_{3a} d\nv^a =
- \frac{\cpl^2}{u^2 + \cpl^2\nvt\hM^\top\hM\nv}
d\nvt \hM \nv,
$$
In particular, at $\mathcal{O}(L)$, we have
$$
    B_{3 a} = - \frac{\cpl^2}{u^{2}} \hM_{ab} \nv^{b}.
$$
When acted upon by $\hO\in SO(6)$, $B_{3a}$ transforms as
$$
   \hO(B)_{3a} =
      -\frac{\cpl^2}{u^{2}} \hO_{ab}\hM_{bc}\hO^{-1}_{cd} \nv^{d}.
$$
Because $\hM$ is in the Lie algebra, $so(6)$, we recognize this as the
adjoint representation, $\rep{15}$
or, in terms of
spherical harmonics, the $k=1, M^{2} = 8$ representation in
\cite{KRvN}. This corresponds to a dimension 5 operator. Referring to
the table of \cite{Intr}, this is a linear combination of
$\delta^{3}\bar\delta\mathcal{O}_{3}$ and
$\bar\delta^{3}\delta\mathcal{O}_{3}$ where $\mathcal{O}_{p}$
is the chiral primary
$\tr{\Phi^{(I_{1}}\Phi^{I_{2}}\cdots\Phi^{I_{p})}}  -
\mathrm{traces}$.
Here, $\delta\mathcal{O}$ ($\bar\delta\mathcal{O}$) represents either
the commutator or
anti-commutator, as appropriate, of $\mathcal{O}$ with the
supersymmetry generator
$Q$ ($\bar{Q}$). We also note from the table that this is a vector
operator, as expected.

For large $u$, we can make the following expansion
$$
     \frac{1}{u^{2} + \a^2} = \frac{1}{u^{2}}\left(1 - \frac{\a^2}{u^{2}} +
      \frac{\a^2}{u^{4}} + \cdots\right).
$$
Thus, we obtain deformations that are multiplied by a higher power of
the dipole length. We will work through some of the
$\mathcal{O}(L^{2})$ terms.

For the deformation of the sphere $\MS{5}$, we have
$$
        h_{(ab)} = - \frac{\mcr^2\cpl^2\hM_{ac}\nv^{c}\hM_{bd}\nv^{d}}{u^{2}}(1 +
        \cO(L^{2})) - \mathrm{trace}
$$
We immediately see from the
above that this must transform in some component of the $\rep{15}
\otimes_\mathrm{sym} \rep{15} = \rep{84} \oplus \rep{20} \oplus
\rep{15} \oplus \rep{1}$. With a little work, one can see that the correct
component is the \rep{84}.
In terms of
spherical harmonics, this corresponds to the $k = 1, M^{2} = 12$ field
in \cite{KRvN}. We can read off the weights from the Young tableaux,
giving $(2,0,2)$. Following the table of \cite{Intr}, we can see that
this corresponds to the operator $\delta^{2}{\bar{\delta}}^{2}
\cO_{4}$. This is a scalar operator of dimension 6.

Also, at $\cO(L^{2})$, there are a number of other deformations arising
from the term $\nvt\hM^\top\hM\nv$. Specifically, there are
$h^{a}_{a}$, the dilaton and the 33 component of the metric on
$\AdS{5}$. As $\hM^{\top}\hM$ is a symmetric $6\times6$ matrix, we can
see that these transform in the $\rep{20} \oplus \rep{1}$.
For the dilaton, we can identify the mass of the
\rep{20} as $M^{2} = 12$. Thus, this is also a dimension 6 operator.
Again, reading from \cite{Intr} identifies the operator as some
combination of $\delta^{4}$ and $\bar\delta^{4}$ acting on $\cO_{4}$.
For the trace of the metric, we are presented with a problem. The metric
has a term in its equation of motion arising from the product of two
three-form NSNS fluxes, giving a term also of order $\cO(L^2)$.
Thus, it can no longer be treated as a linear
perturbation on the $\AdS{}$ background. The fact that we got the correct
answer for the traceless part of the metric perturbation is due to the fact
that the product of the NSNS fluxes does not have any component that
transforms in the {\bf 84}, and so it can be treated as a linear
perturbation.

\section{The Geometry of the Supergravity}\label{geom}

We now investigate some of the geometrical features of the metric
(\ref{twmet}).
The key things to note are the behavior of $x_3$ coordinate and
the $\MS{5}$ as a function of $u$.
  We first discuss the generic behavior and then give a detailed analysis
of a useful special case that will occupy us for the remainder of the
paper.
General deformations of the $\MS{5}$ were also studied in a slightly
different context in \cite{FGPW}.

\subsection{The Boundary}

The behavior near the boundary is governed by the rank of $\hM$.
For maximal rank, the quadratic function $\nvt \hM^{\top}\hM \nv$
is always positive definite. It has 12 local extrema on $\MS{5}$.
These consist of pairs of antipodal points -- each pair corresponds
to an eigenvector of $\hM^{\top}\hM$ with the two ($\pm$)
sign options.
The metric (\ref{twmet})
    is asymptotically $\AdS{4}\times \MS{1} \times \MS{5}$ where the
$\MS{5}$ is deformed, and both the $\MS{1}$ and $\MS{5}$ are small
compared to the $\AdS{}$.

If the rank of $\hM$ is less than maximal, the quadratic
form $\nvt\hM^{\top}\hM\nv$ has a locus of zeroes.
This locus is $\MS{r-1}$ where $r=2,4$ are the possible nonzero
values for the rank.
Locally on the zero
locus, the metric is indistinguishable from ordinary $\AdS{5} \times
\MS{5}$. This should be related to the fact that some scalar
fields do not have a dipole length.
We do not claim to understand the exact connection.

The metric on $\MS{5}$ becomes degenerate as $u\rightarrow 0$.
For $\hM$ of maximal rank, the metric on $\MS{5}$ at $u=0$ is
$$
ds^2 =  \mcr^2 d\nvt d\nv -
        \frac{\mcr^2(\nvt \hM^\top d\nv)^{2}}
        {\nvt \hM^\top\hM \nv}
$$
Let $\nv$, a unit vector in $\MR{6}$, parameterize a point $p\in\MS{5}$.
Then $\hM\nv$ defines a direction in the tangent space $T_p\MS{5}$, since
$\nvt\hM\nv = 0$. It is easy to see that the metric is degenerate along
this direction. Thus, $\hM\nv$ defines a vector field on $\MS{5}$ along which
the metric is degenerate. This is the vector field induced by the
infinitesimal $SO(6)$ action on $\MS{5}$ given by $\hM \in so(6)$.
To analyze the degenerate $\MS{5}$ further we need to know more about
the eigenvalues of $\hM$.
Let the eigenvalues be $\pm i\a_1,\pm i\a_2,\pm i\a_3$.
If $\a_1=\a_2=\a_3$ then the flow lines of the vector field $\hM\nv$ are
closed circles. $\MS{5}$ can be described as a circle bundle over
$\CP{2}$, and the vector field is along the circle. At $u=0$ the
$\MS{5}$ then shrinks to $\CP{2}$.
This particular case will be discussed more extensively in
the next section. For the general case, we can identify $\MR{6}$
with $\MC{3}$ and introduce the following coordinates
$$
(z_1,z_2,z_3)=\left(\frac{e^{i\a} r\cos\th}{\sqrt{1 + r^{2}}},
\frac{e^{i\b}r\sin\th}{\sqrt{1 + r^{2}}},\frac{e^{i\g}}{\sqrt{1 +
r^{2}}}\right)
$$
In these coordinates, the deformation of the sphere only affects the
three coordinates $\a$, $\b$ and $\g$. The vector $\hM \nv$ is solely
along this torus and, for generic ratios between these angles, the flow
is dense in this torus. However, this is not a true fibration, and
to avoid such complications we will only work with the simpler case.

\subsection{The Hopf Fibration}\label{spxmpl}
The case when all three eigenvalues of $\hM$ are equal is the case
where all of the scalar fields have the same dipole length. The
analysis of the UV behavior of the theory will significantly simplify
in this situation.

We set $\frac{1}{2}\dipL\equiv\a_1=\a_2=\a_3$ and
\be\label{spform}
\hM = \left(\begin{array}{rrrrrr}
                0&-\dipL& 0& 0& 0& 0\\
                \dipL& 0& 0& 0& 0& 0\\
                0& 0& 0&-\dipL& 0& 0\\
                0& 0& \dipL& 0& 0& 0\\
                0& 0& 0& 0& 0&-\dipL\\
                0& 0& 0& 0& \dipL& 0\\
\end{array}\right).
\ee
For $\dipL\neq 0$,
this choice of $\hM$ breaks all of the supersymmetry but it preserves
a $U(3)\subset SO(6)$ subgroup of the R-symmetry.
The advantage of this choice of $\hM$ is that the factor
$\nvt\hM^\top\hM\nv =\dipL^2$  is independent of $\nv$.
According to the definition of $\hM$ in subsection (\ref{symdef}),
the bosons have dipole lengths $\pm 2\pi\dipL$, three of the fermions
have dipole lengths $\pm\pi\dipL$ and the remaining (complex) fermion has
length $\pm 3\pi\dipL$.

We now write the metric on the deformed $\MS{5}$ explicitly.
Let a unit vector $\nv$ which parametrizes
$\MS{5}$ in $\MC{3}$ be given by
$$
\nv = \left(
\frac{e^{i\gamma}}{\sqrt{1+|\alpha|^2+|\beta|^2}},
\frac{e^{i\gamma}\alpha}{\sqrt{1+|\alpha|^2+|\beta|^2}},
\frac{e^{i\gamma}\beta}{\sqrt{1+|\alpha|^2+|\beta|^2}}
\right)
$$
with $\a$ and $\b$ complex. Thus, the $\MS{5}$ is given as a circle
fibration parametrized by $\gamma$ over $\CP{2}$ parametrized by
$\alpha$ and $\beta$. This is the famed Hopf fibration. The advantage
of these coordinates is that the vector $\hM \nv$ points along the
direction of the fiber for $\hM$ given as in (\ref{spform}).

It can be shown that the metric on a regular $\MS{5}$ is
in these coordinates
\bear
d\nvt d\nv &=&
\frac{|d\alpha|^2+|d\beta|^2}{1+|\alpha|^2+|\beta|^2}
-\frac{|\bar\alpha d\alpha + \bar \beta d\beta|^2}
{(1+|\alpha|^2+|\beta|^2)^2}
+\left (
d \gamma +
\frac{\Im(\bar\alpha d\alpha + \bar \beta d\beta)}
{1+|\alpha|^2+|\beta|^2}
\right )^2
\label{cpabg}
\eear
where the first 2 terms describe the Fubini-Study metric on $\CP{2}$.

For our deformed sphere, the metric is
\bear
d\nvt d\nv - \frac{\cpl^2(\nvt\hM^\top d\nv)^2}{u^2+\cpl^2\nvt \hM^\top \hM \nv} &=&
\frac{|d\alpha|^2+|d\beta|^2}{1+|\alpha|^2+|\beta|^2}
-\frac{|\bar\alpha d\alpha + \bar \beta d\beta|^2}
{(1+|\alpha|^2+|\beta|^2)^2}
\nn\\
&&+ \frac{u^2}{u^2+\cpl^2\dipL^2}
\left (
d \gamma +
\frac{\Im(\bar\alpha d\alpha + \bar \beta d\beta)}
{1+|\alpha|^2+|\beta|^2}
\right )^2
\nn
\eear

The $5\times 5$ determinant of the above metric
can be calculated to be
\begin{equation}
\label{spheredet}
\mathrm{det\ }g = \left(\frac{u^2}{u^2+\cpl^2\dipL^2}\right)
\frac{1}{(1+|\alpha|^2+|\beta|^2)^6}
\end{equation}

Thus, the salient features of our deformed $\MS{5}$ are as follows
\begin{itemize}
\item
It has the structure of an $\MS{1}$
(Hopf) fibration over a base $\CP{2}$.
An $SU(3)$ subgroup of $SO(6)$ acts freely on $\CP{2}$.
\item
Invariance of the
metric of the deformed $\MS{5}$ under $U(3)\subset SO(6)$ implies
that the metric on the base $\CP{2}$ is independent
of the position, and the metric on the fiber $\MS{1}$ is similarly homogeneous
due to the $U(1)$ isometry which rotates the fibers.
\item
The radius of the fiber is independent of the $\CP{2}$
coordinate and is given by
\be\label{rhodef}
\rho(u) = \mcr\frac{u}{\sqrt{u^2+\cpl^2\dipL^2}}.
\ee
\item
The volume of $\CP{2}$ is constant and given by
$$
{\mbox{Volume}}(\CP{2}) = \frac{\pi^2}{2}\mcr^4.
$$
\end{itemize}

Finally, in these coordinates, the NSNS 2-form is given by
$$
B =
- \frac{\cpl^2\dipL}{u^2 + \cpl^2\dipL^2} dx_3\wedge \psi.
$$
Here $\psi$ is the global angular 1-form of the Hopf fibration.
In the notation of (\ref{cpabg}), it is given by
$$
\psi =
d \gamma +
\frac{\Im(\bar\alpha d\alpha + \bar \beta d\beta)}{1+|\alpha|^2+|\beta|^2}.
$$
The 3-form field strength is given by
$$
H = dB =
- \frac{\cpl^2\dipL}{u^2+\cpl^2\dipL^2} dx_3 \wedge d\psi
+ \frac{\cpl^2\dipL u}{(u^2+\cpl^2\dipL^2)^2}du\wedge dx_3\wedge \psi.
$$
Here $d\psi$ is the closed harmonic 2-form that generates
$H^2(\CP{2},\BZ)$.

\section{Nonlocality in the Supergravity Dual}\label{nonloc}
We now come to the heart of the paper. In this section, we will show
how the nonlocality of the field theory is manifested in the geometry
of the boundary of the supergravity. We will continue to work with the
special case described above in (\ref{spxmpl}).
In this situation, as described in that section,
the fiber shrinks to zero size on the boundary, and, as such,
should be T-dualized to obtain a classical description. This
will make the dipole nature of the nonlocality evident.

\subsection{T-duality of the Fiber}\label{xtdl}
As we approach the boundary of our solution, $u\rightarrow 0$,
the volume of the base $\CP{2}$ remains a constant.
However, the circle fibered along it shrinks to zero size.
Note that the dilaton also approaches zero since
$$
e^{2(\phi-\phi_{0})} = \frac{u^2}{u^2 + \cpl^2\dipL^2}
$$
It is easy to see that the curvature of the deformed $\MS{5}$
is still of the order of magnitude of $\frac{1}{\mcr^2}$, even
when $u\ll \cpl\dipL$. However, when $\rho(u)$ becomes of the order of
magnitude of the string length, $\a'^{1/2}$, we cannot trust
the supergravity approximation anymore.
This happens when $u\sim \a'^{1/2}\mcr^{-1}\cpl\dipL = \cpl^{1/2}\dipL$.

Since the circle shrinks to zero, we have to perform T-duality on that
direction. As we shall see in section (\ref{fermions}), there is a subtlety
that complicates matters, but for the time being we will naively apply the
standard T-duality formulae.

Again using the equations of \cite{BUSH,BUSC,BHO},
we obtain type-IIA
with the metric
\bear
      ds^{2} &=& \frac{\mcr^2}{u^2}
     (dt^{2} - dx_{1}^{2} - dx_{2}^{2} -du^2)
        - \frac{\mcr^2}{u^2} (dx_3 + \dipL d\gamma)^2
     -\frac{{\a'}^2}{\mcr^2} d\gamma^2 \nn\\
     & & \qquad\qquad- ({\mbox{constant $\CP{2}$}}).
\label{mettdl}
\eear

We also have
\bear
e^{2\phi} &=& \frac{e^{2\phi_0}}{\cpl} = \sqrt{\frac{g_s^3}{4\pi N}},
\nn\\
\sum_b H_{u3b}dx^b &=& - \frac{\cpl^2\dipL u}{(u^2+\cpl^2\dipL^2)^2}
\frac{\Im(\bar\alpha d\alpha + \bar \beta d\beta)}{1+|\alpha|^2+|\beta|^2}
.\nn
\eear
where $H$ is the 3-form NSNS field strength.
In addition, there is a non-trivial RR 4-form field strength which we will
not write down.
Note that the type-IIA dilaton becomes a constant.
Despite the ominous factor $\frac{{\a'}^2}{\mcr^2}\ll 1$
in (\ref{mettdl}), we see that
type-IIA supergravity is a good approximation.  No two points
that are closer than ${\a'}^{1/2}$ are identified.
The only identification is
$$
(\dots,x_3,\gamma)\sim (\dots,x_3,\gamma+2\pi)
$$
and the distance between those two points is large when $u\rightarrow 0$.

\subsection{Nonlocality on the Boundary}\label{nlcbdry}

The metric in equation (\ref{mettdl}) is a striking manifestation of
the nonlocality of the field theory in the boundary metric.
It describes the $x_3$ direction fibered over a small circle of radius
$\frac{{\a'}}{\mcr}$ parameterized by $\gamma$.
The proper distance between the
point with coordinates $(x_3,\gamma)$ and the point with
coordinates $(x_3 + 2\pi\dipL, \gamma)\sim (x_3,\gamma-2\pi)$
is $\frac{2\pi\a'}{\mcr}$ which is of stringy
scale. On the other hand, the proper distance between $(x_3,\gamma)$ and
$(x_3 + \Delta,\gamma)$ is of order $\frac{\mcr}{u}\rightarrow \infty$
when $\Delta$ is not an integer multiple of $2\pi\dipL$
and $u\rightarrow 0$.
In the field theory, this is translated into nonlocal interactions
between fields at points that are separated by a distance of 
$L=2\pi\dipL$.
If we think of the matter content of the dual SYM theory as constituting
momentum modes along the $\MS{5}$, then, after T-duality, the
nonlocality should be reflected in the winding number around the
T-dual circle. This is exactly what we see here.

The metric (\ref{mettdl})
also shows that the 4D superconformal group is restored since
the new coordinate $x_3+\dipL\gamma$ can be attached to the
$\AdS{4}$ part of the metric to form $\AdS{5}$. This is to be expected
because the nonlocal interactions have a minimal distance $L$.
At short distances the vicinity of each point should look like a 4D CFT
and the interactions with fields at distance
$L$ seems like an interaction
with extra degrees of freedom outside the small neighborhood of the point.

\subsection{A Note on Momentum Conservation}

It is interesting to note that, because $\MS{5}$ is contractible, the
winding number along the $\MS{1}$ fiber is not conserved. This is
equivalent to the fact that the fibration has a nontrivial first
Chern class. In order to contract the circle, however,
one needs to pull it around a nontrivial 2-cycle of the base $\CP{2}$.
So, a concrete process that violates winding number conservation
is to start with a small string on $\CP{2}$ and then to gradually increase
its size until it extends around the equator of a topologically
nontrivial $\CP{1} \cong \MS{2}$ inside  $\CP{2}$.
Then we contract the string along the
other hemisphere of the $\CP{1}$.
At the end of the process, the string is wound around the fiber $\MS{1}$.
This process requires energy scales of the order of
the circumference of the equator of the $\CP{1}$, {\it i.e.\/},
$E\sim \mcr / \a' $

Because of this, after T-duality, momentum along the
$\gamma$-direction  also must not
be conserved. After T-duality, the $\gamma$-circle is fibered trivially
over the $\CP{2}$. Instead,
we have a 3-form NSNS field strength,
$H_{\gamma ab}$, along the circle and two directions inside the $\CP{2}$.
It is easy to see that $H_{\gamma ab}$ is proportional to
$d\gamma \wdg \omega$ where
$\omega$ is the harmonic 2-form on $\CP{2}$.

The process that violates momentum conservation along the
$\gamma$-direction is the same as before.
We start with a pointlike string inside
$\CP{2}$ and deform it to go around a nontrivial 2-cycle inside
$\CP{2}$ and then shrink it back to a point.
Let $X(\sigma,\tau)$ be the closed path of the string as a function
of time $\tau$ and string coordinate $0\le \sigma\le 2\pi$.
Note that when both $\sigma$ and $\tau$ vary, the function
$X(\sigma,\tau)$ spans a surface that is homologically equivalent
to the nontrivial 2-cycle inside $\CP{2}$.
The violation of momentum conservation is due to the ``magnetic''
forces on a moving string in the presence of an $H=dB$ field strength.
The total $\gamma$-momentum transfer is
$$
\int F_\gamma(\tau)d\tau = \int H_{\gamma ab} \px{\sigma}X^a\px{\tau}X^b
     d\sigma d\tau =\int \omega = 1.
$$
The RHS is the integral of the 2-form $\omega$ along the nontrivial
2-cycle.

\section{The Fermions}\label{fermions}
In the previous section we saw that T-duality on the $\MS{1}$ fiber
of the deformed $\MS{5}$ leads to a simple picture of nonlocality
on the boundary.
The nonlocality scale, $L=2\pi\dipL$, of the field theory
matched nicely with the nonlocality scale on the boundary.
In general, the proper distance between any two distinct points
along the $x_3$-axis becomes infinite on the boundary because of the
large rescaling factor $\frac{1}{u^2}$. If the $x_3$ coordinates
of the two points differ by an integer multiple of $L$ then, 
as we saw in subsection (\ref{nlcbdry}),
one can make a ``shortcut'' through an extra dimension that came from 
the T-dual of the $\MS{1}$ and go from one point to the other 
via a path whose proper length is shorter than the string scale.

However, the logic behind this picture is incomplete.
To understand the problem, we will begin with a puzzle.

\subsection{What About the Fermions?}
The supergravity metric presents a nonlocal behavior that connects
two points at $x_3$-distance of $L$, and this is indeed the dipole vector
of the scalars of our field theory. But what about the fermions?
Their dipole vectors, as mentioned below equation (\ref{spform}), are
$\pm\frac{L}{2}$ or $\pm\frac{3L}{2}$.

One might try to argue that we should only consider fermion bilinear
operators but this does not appear to be the case.
Obviously, there are fermionic operators in the theory.
Moreover, let us consider gauge invariant operators in the field
theory that also carry R-symmetry charge.
Specifically, let us consider the $U(1)_c$ center of the
$U(3)\subset \Spin{6}_R$ that keeps the dipole matrix
(\ref{spform}) invariant.
This $U(1)_c$ acts on the $\MS{5}$, and it is easy to see that
it is represented by rotations of the fiber $\MS{1}$.
The dipole vector of any field $\Phi$ is given by $\frac{1}{2}L$
times its $U(1)_c$  charge. To make a gauge invariant operator
we need to include an open Wilson line, for example
\be\label{Wop}
W = \tr{\Phi(x) e^{i\int_C A_\u dx^\u}},
\ee
where $C$ is an open path whose endpoints are at $x_3-\frac{L}{2}$
and $x_3+\frac{L}{2}$.

In the supergravity dual, closed Wilson lines correspond
to closed paths on the boundary \cite{MalW}.
The operator $W$ will also correspond to a closed path.
It is the path that starts along $C$ on the boundary and
then winds around the T-dual $\MS{1}$ to make the shortcut from
$(x_3-\frac{L}{2})$ to $(x_3+\frac{L}{2})$.
Note that after T-duality, the $U(1)_c$ charge is mapped to
winding number along the T-dual $\MS{1}$.
See also \cite{GHI} for a related discussion.

Now suppose that $\Phi$ is a fermion dipole field of length
$\frac{L}{2}$. There is no way to close the Wilson line
in the supergravity dual.

\subsection{A Missing $(-)^F$}
The problem with the T-duality argument of subsection (\ref{xtdl})
is revealed by a careful analysis of the boundary conditions
of the fermions around the fiber $\MS{1}$ of the Hopf fibration
of $\MS{5}$. As we will now argue, the fermions have anti-periodic
boundary conditions around the $\MS{1}$, and one should include
$(-)^F$ in the boundary conditions, where $F$ is the fermion number.
Our setting is reminiscent of the geometry in \cite{WitBH}.

An observer living on our deformed $\MS{5}$ who cannot venture out
over distances of the order of $\mcr$ sees only a small neighborhood
$\ngb$ of $\CP{2}$. In this neighborhood, fields vary slowly
and the fiber $\MS{1}$ is noncontractible. The fibration has
the structure of $\ngb\times\MS{1}$, and, if our local observers wish
to describe fermions in their neighborhood, they have the option of
choosing either periodic or anti-periodic boundary conditions
around $\MS{1}$. The geometric holonomy around $\MS{1}$,
calculated from the Levi-Civita connection, is the identity 
in $SO(5)$.
However, as it turns out, the small $\MS{1}$ fiber
is contractible inside the whole deformed $\MS{5}$,
but in order to shrink it to a point one must first deform the
circle to a path of length at least $2\pi\mcr$.
This fact allows one to calculate the holonomy for fermions
around the fiber $\MS{1}$, and, as we will see below, it is
$-1\in\Spin{5}$.
Thus a local observer would have to choose the anti-periodic boundary
conditions and insert $(-)^F$ in every calculation.

In more mathematical terms,
let $\mathbf{T}_*\MS{5}$ be the tangent bundle over
$\MS{5}$. To define spinors on $\MS{5}$ we need the spin bundle,
$\sbd$ over $\MS{5}$. The structure group of $\sbd$ is 
$\Spin{5}$ which is a double cover of $SO(5)$. Now pick a fiber
$\MS{1}$ over a point $p$ of the base $\CP{2}$. Take a neigborhood
$\ngb\in\CP{2}$ of the point $p$. 
The restriction of the $\MS{1}$-fibration 
to  $\ngb$ is a manifold that is of the form $\ngb\times \MS{1}$.
Over $\ngb\times\MS{1}$ there are two possible spin structures.
In the first one, $\sbd_{+}$,
the spinors have periodic boundary conditions
around the $\MS{1}$ and in the second, $\sbd_{-}$,
the spinors have anti periodic boundary conditions.
The appropriate spin structure can be calculated from the $\Spin{5}$
holonomy around $\MS{1}$ in $\sbd$.
It turns our that the holonomy is $-1\in\Spin{5}$.
To see this one can continuously deform the fiber $\MS{1}$
to a point inside $\MS{5}$ and trace the holonomy around the 
closed loop as it changes from $1\in\Spin{5}$ when the loop is a point
to $-1\in\Spin{5}$ when the loop becomes the fiber.
To actually calculate the holonomy, note that
the fiber $\MS{1}$ is a circle of radius $\mcr$ inside $\MS{5}$.
Pick an $\MS{2}\subset \MS{5}$ that contains $\MS{1}$ as its equator.
Over $\MS{2}$, 
$\sbd$ reduces to the spin bundle of $\MS{2}$ times a trivial bundle.
So the holonomy is the same as the holonomy of a spinor on $\MS{2}$
around the equator which is $-1$.

It may seem at first sight that because of the $(-)^F$
we are actually describing strings at high temperature
as in \cite{AtiWit} (and see also 
\cite{GGKRW}-\cite{ABKR} for recent discussions).
The closed string spectrum would then develop a tachyon
when the $\MS{1}$ shrinks to a size smaller than the string scale,
and our discussion would be rendered invalid.
However, the same $-1\in\Spin{5}$ holonomy is there even for
the supersymmetric $\AdS{5}\times\MS{5}$ since the geometry of the
$\MS{5}$ is the same except for the size of the fiber.
This means that the $\MS{5}$ must support covariantly constant 
spinors, and something else should cancel the $-1\in\Spin{5}$ phase.
Indeed, the Dirac equation of motion for a fermion $\psi$ on 
$\AdS{5}\times\MS{5}$ contains an extra term, in addition to
the spin connection. This term is proportional to 
$(F_5)_{\u_1\dots\u_5}\Gamma^{\u_1\dots\u_5}\psi$ where
$F_5$ is the 5-form RR field strength (see, for example, \cite{KleWit}).
When integrated around the fiber $\MS{1}$, this extra term
gives an additional phase of $(-1)$ so that altogether a covariantly
constant spinor is possible for an $SO(6)$-symmetric $\MS{5}$.

It is important to point out that the $(-)^F$ phase coming
from the geometric holonomy is a global effect.
If we return to our local observer on $\MS{5}$,
the $(-)^F$ rule will seem to them as an arbitrary rule of 
nature. On the other hand the $(-)^F$ phase coming from $F_5$
can be calculated locally. This has important implications
to the application of T-duality.

\subsection{T-duality with $(-)^F$}
The anti-periodic boundary conditions for the fermions around 
the fiber $\MS{1}$ imply that we cannot just perform T-duality
and get a type-IIA background with a large $\MS{1}$.
Instead we get a type-0A theory.
Such theories where discussed in 
\cite{DixHar}-\cite{KleTse}.
Their spectrum contains no fermions, and their bosonic massless spectrum 
is the same as that of type-IIA string theory but with two copies of
every field in the RR sector.
The main complication is that they also contain a tachyon.
However, in our case the tachyon could very well be absent.
In \cite{AtiWit} the tachyon came from a string winding state
in the RR sector. There, because of the extra $(-)^F$, there was
a negative zero point energy for the worldsheet oscillators.
In our case, as we have seen above, the $F_5$ term cancels the
$(-)^F$, and therefore the winding state is quite likely to remain 
massive.  The disappearance of the tachyonic instability
is also supported by the arguments of \cite{KleTse}.
There it was argued that a background RR flux provides a positive
shift to the $({\mbox{mass}})^2$ of the tachyon.
If that is indeed the case, it is plausible that the UV region
is described by type-0A string theory with the general features 
of the weakly curved metric described in section (\ref{xtdl}).
Note that the magnitude of the 4-form RR field strength in type-0A
is $\frac{M_s^4}{g_s}=g_s^{1/3} M_p^4$. This means that when $g_s$
is small, this field strength is large relative to the string scale 
but small relative to the Planck scale.
Thus it appears that our dipole theory describes an RG flow from type-IIB
string theory to type-OA string theory.\footnote{We are grateful
to Igor Klebanov for pointing this out.}
However, quantizing strings in strong RR backgrounds remains an open problem,
and type-0A is also likely to have a large cosmological constant, so this
conjecture is hard to verify.

\subsection{Resolution of the Puzzle}
Assuming that T-duality to a weakly coupled type-0A theory is
possible, the puzzle about spinor operators with half-integral
dipole length is resolved as follows.
The type-IIB compactification on $\MS{1}$ of radius $R$ with the
extra $(-)^F$ twist can be described as the orbifold of
a compactification on a circle of radius $2R$ by the $\BZ_2$
action $(-)^{F+P}$ where $P$ is the Kaluza-Klein momentum.
The T-dual is therefore an orbifold of type-IIA on a circle of
radius $\frac{1}{2R}$ by $(-)^{F+W}$ where $W$ is the winding number.
Now we see that in the untwisted sector of the
T-dual background, strings that are spacetime
bosons must have even winding number, and strings that are spacetime
fermions must have odd winding number.
Thus, the resolution of the puzzle is that the dual theory 
is not type-IIA on a circle of radius $\frac{1}{R}$ but rather type-0A
on a circle of radius $\frac{1}{2R}$.

\section{Correlation Functions}\label{corrfcn}

In a local field theory, correlation functions of operators,
$\ev{O(x) O(y)}$, have short distance singularities when $x\rightarrow y$.
In dipole theories, we expect a singularity to appear also when
$x\rightarrow y\pm L_i$, where $L_i$ is one of the characteristic vectors
of nonlocality as in section (\ref{DipTh}). In the special case we study
in this paper, the length of the characteristic vectors of the scalars is
$L=2\pi\dipL$.
For operators $O(x)$ that have no dipole length of their own
(for example $\tr{F_{\u\v}^2}$) we therefore expect
$$
\ev{O(x) O(y)}\longrightarrow_{x\rightarrow y+L}
    \frac{C}{|x-y-L|^{2\Delta}}
$$
and then in momentum space we expect  to find a term that behaves like
$$
\ev{O(k)^\dagger O(k)} \longrightarrow_{k\rightarrow\infty}
\frac{C e^{i k\cdot L}}{k^{4-2\Delta}}.
$$
For operators $O(x)$ that do have a length we expect the behavior 
of the correlation function to be
more complicated since the operators contain nonlocal Wilson lines
as in (\ref{Wop}). It is likely that the correlation functions
exhibit an exponential behavior $\sim e^{\sqrt{({\mbox{const}}) |k_3| L}}$
analogous to that of noncommutative geometry \cite{GHI,vRR}.\footnote{We
are grateful to M.~Rozali for a discussion on this point.}
For the rest of this discussion we will restrict ourselves
to operators $O(x)$ with dipole length zero.

We can use the AdS/CFT correspondence to compute these correlation
functions in the large $N$ limit.
We will restrict ourselves to the special case where the R-symmetry
is broken from $\Spin{6}$ down to $U(3)$, as in section (\ref{spxmpl}).
Because the AdS/CFT correspondence directly probes the
nonperturbative nature of the field theory, it is perhaps a bit too much to
expect to see the exact form above, but, in the limit of high momentum
along the dipole direction, a sign of nonlocality would be a rapid
oscillation in the correlation function in momentum space.

It is, in general, a difficult problem to decouple the fields on a
nontrivial background such as any of the examples in this paper.
Following \cite{GHI}, we will
simply postulate that there exists a massless scalar living on our
spacetime.\footnote{We are grateful to I.R. Klebanov
for explaining the relevant issues to us.}
    In particular, it should satisfy the field equation
$$
        \partial_{\mu}\left(e^{-2\phi} \sqrt{\mathrm{det\ }g}\ g^{\mu\nu}
        \partial_{\nu} \Phi(\xv,u)\right) = 0
$$
where $\xv = (t,x_{1},x_{2},x_{3})$.

This is still quite a difficult problem to solve, but we will soon
see how it can be simplified. In particular, we recall the
determinant of the metric of the sphere, (\ref{spheredet}), in the Hopf
fibration coordinates.
Including the $\AdS{}$ portion of the metric, we have
$$
        \mathrm{det\ }g = \mcr^{20} \left(\frac{u^{2}}{u^{2} +
  \cpl^2\dipL^{2}}\right)^{2}
        u^{-10} \frac{1}{(1 + |\alpha|^{2} + |\beta|^{2})^6}
$$
We immediately see that this factors into a contribution that depends
on the sphere and one that depends on the $\AdS{}$.
Thus, because our metric
is block diagonal, we can choose our scalar field to be constant on
the sphere, and all contributions from the sphere will cancel out of
our equations.
Another happy fact is that the contribution from the dilaton exactly
cancels the $u^{2}/(u^{2} + \cpl^2\dipL^{2})$
term reducing this to almost the
standard massless field equation on $\AdS{}$ space.

As usual, the most interesting part of the equation comes from the
$u$ coordinate, so we write
$$
\Phi(\xv,u) = \varphi(u) e^{i \kv \cdot \xv}
$$

Then $\varphi$ satisfies the following equation
$$
 u^{3} \partial_{u} \left(\frac{1}{u^3} \partial_{u} \varphi(u)\right) 
+\left(k^{2} - \frac{(\cpl k_{3} \dipL)^{2}}{u^{2}}\right)\varphi(u) 
  = 0
$$
If we expand this, we obtain
$$
        \varphi'' - \frac{3}{u} \varphi' + \left(k^{2} -
        \frac{(\cpl k_{3} \dipL)^{2}}{u^{2}}\right)\varphi = 0.
$$
We recognize this as the equation for a massive field in ordinary
$\AdS{}$ space with $m \mcr = \cpl k_{3} \dipL$.
Thus, we can copy the final result from equation (44) of \cite{GKP}
$$
        \langle\mco(k)\mco(q)\rangle = -(2\pi)^{4}\delta^{4}(k + q)
        \frac{N^{2}}{8\pi^{2}} \frac{\Gamma(1-\nu)}{\Gamma(\nu)}
        \left(\frac{k \mcr}{2}\right)^{2\nu} \mcr^{-4}
$$
where $\nu = \sqrt{4 + (k_{3} \dipL)^{2}}$.

Let us now take the limit that $k_{3} \to \infty$. In this
limit, we have
$$
\langle\mco(k)\mco(-k)\rangle
\sim \frac{1}{\sin(\pi \nu)}
\left (
\frac{(k\mcr/2)^{\nu}}{\Gamma(\nu)}
\right)^2
\stackrel{k_{3}\to\infty}{\sim}
\frac{\cpl k_{3}\dipL \left(\frac{|k| e \mcr}{2\cpl k_{3}\dipL}
\right)^{2 \cpl k_{3}\dipL}}
{\sin(\pi \cpl k_{3} \dipL)}
$$
It exhibits an oscillatory behavior but not quite what we have 
anticipated.
We  expected the wavelength of the nonlocal
behavior to be an integer multiple of the dipole length. This is not
what we observe here. 
This is a puzzling phenomenon, but it is consistent
with the observation from the supergravity dual that the scale of the
nonlocality is actually $\cpl \dipL$ rather than just $\dipL$. 
Since $\cpl\dipL >>\dipL$ there is no immediate contradiction.
It could be that in the large $\cpl$ limit the dominant 
contribution to the nonlocal behavior of the correlation function
comes from the nonlocality on scale $[\cpl]\dipL$ (where $[\cpl]$
is the integer that is closest to $\cpl$).
It is important to realize, however,
that the supergravity approximation ceases to be valid
when $u< {\a'}^{1/2}\mcr^{-1}$, as we explained in subsection (\ref{xtdl}).
This suggests that the above calculation may not be entirely valid.
This is worthy of further investigation.

\section{Discussion}\label{disc}

In this paper, we have shown how the nonlocality of dipole theories is
manifested in the supergravity dual. We discovered that the metric
becomes degenerate at the boundary of the spacetime and that this could
be used to explicitly demonstrate the nonlocality. 
Although this feature of the metric was shown using the naive T-duality
to type-IIA and, as we argued in section (\ref{fermions}), one actually
gets type-0A with a strong RR field strength, we believe that the metric
still has this general structure.
This should be
a generic feature of the supergravity duals of nonlocal field
theories. It is not a surprising result. Nonlocality, when realized in
some limit of string theory, cannot be a purely supergravity effect.
The nonlocality must be a result of the inclusion of some stringy
degrees of freedom on the boundary. The degeneracy of the metric in
string frame means that we cannot treat the boundary as classical, and
this is the source of the nonlocality.

It is worthwhile to compare this situation to that in noncommutative
geometry to see if we can distill some more general features of the
supergravity dual. 
The discussion that follows has some features in common
 with \cite{DH},\cite{HI}-\cite{Sara}.
\footnote{We are grateful to A. Hashimoto for pointing out some of these
references and for discussing this with us.}

Recall that the metric of the supergravity dual of
NCSYM is \cite{HasItz,MalRus} (ignoring dimensionless constants)
$$
ds^{2} =\frac{1}{u^{2}}\left(dt^{2} - dx_{1}^{2} - \frac{u^{4}}
{u^{4} + \theta^{2}}\left(dx_{2}^{2} + dx_{3}^{2}\right) -
du^{2}\right)
$$
The other fields are
\begin{eqnarray*}
       e^{2\phi} &=& \frac{u^{4}}{u^{4} + \theta^{2}} \\
       B_{23} &=& -\frac{\theta}{u^{4} + \theta^{2}}
\end{eqnarray*}

We see that both the second and third directions go to zero length on
the boundary, indicating some sort of stringy effect. Note that here,
the degeneracy is in the $\AdS{}$ part of the metric indicating that the
nonlocality is part of the space that the field theory lives on. This is in
contrast to our dipole theories where the degeneracy is on the
$\MS{5}$ indicating that the nonlocality is part of the field content
of the theory.

Following the
same procedure as in (\ref{nlcbdry}), we compactify these directions and
T-dualize along one of them,
say the second.
As before, the presence of the B-field gives rise to cross
terms in the metric. Specifically, after T-duality, we have,
isolating the 2 and 3 directions
$$
ds^{2} = (u^{4} + \th^{2})\frac{dx_{2}^{2}}{u^{2}} +
2\th\frac{dx_{2}dx_{3}}{u^{2}} +
\frac{u^{2} + \th^{2}u^{-2}}{u^{4} + \th^{2}}dx_{3}^{2}
$$

If we take the $u\to 0$ limit, we can rewrite this as
$$
ds^{2} = \frac{1}{u^{2}}(\th dx_{2} + dx_{3})^{2}
$$

This has almost the same form as the metric we obtained in section
(\ref{xtdl}). When we traverse the 2-circle, the above
coordinate gets shifted by $\th$. As T-duality interchanges momentum
with winding, we interpret this as a dipole in the 3 direction with
length equal to $\th$ times the momentum. This is exactly the
situation in NCSYM.

What are the general features of the supergravity duals of
nonlocal field theories that we can infer from this?
\begin{itemize}
       \item{The metric becomes degenerate on the boundary of $\AdS{}$.}
       \item{The NSNS 2-form field has a component along the
       degenerate direction.}
       \item{We can (perhaps after compactification) T-dualize along this
       direction.}
       \item{After T-duality, the NSNS 2-form field induces off-diagonal
       terms in the metric that can be interpreted as a fibration over
       a string scale circle.}
       \item{The nonlocality of the field theory is manifested by the
       shift in the new coordinate as we go around the string scale circle.}
\end{itemize}
While these features may not be generic for all nonlocal theories, it is
not unreasonable to assume that they may be generic for the
generalized dipole theories mentioned at the end of section
(\ref{dipdef}) of which both the dipoles discussed here and those of
noncommutative geometry are a special case. In \cite{BerGan} a
generalization of dipoles to the case of the (2,0) theory was proposed
where, instead of constant length dipoles, there are constant area
``discpoles''. This should
have a supergravity dual of the from $\AdS{7} \times \MS{4}$. It
would be interesting to investigate the effects of nonlocality on the
supergravity in this situation.

Before concluding let us return to a loose end from the beginning
of section (\ref{strth}).
We mentioned that the twisted string theory backgrounds are unstable
if supersymmetry is broken. This instability was discussed in
\cite{DGGH,CosGut,Str} and is related to the instability of 
Kaluza-Klein compactifications without supersymmetry \cite{WitKK}.
In section (\ref{spxmpl}) we used a nonsupersymmetric 
twisted theory, and we therefore expect it to be unstable.
However, the probability for decay per unit time and volume is 
exponentially suppressed as $g_s\rightarrow 0$. In the large $N$ limit 
(keeping $g_s N$ fixed) we can therefore assume that the background is stable.
It is interesting to ask whether the dipole field theory on the probe is
also unstable. We will not address this question here. One possibility suggested
by O. Aharony is that a potential is generated on the Coulomb branch of the
dipole field theory that makes the origin unstable.
This is currently under investigation.


\section*{Acknowledgments}
We are grateful to O. Aharony, J. Gomis, R. Gopakumar, A. Hashimoto,
C.P. Herzog, N. Itzhaki, I.R. Klebanov, J. Maldacena, D. Minic, S. Minwalla,
B. Morariu, A. Polychronakos, N. Seiberg,
S. Shatashvili, S. Shenker, A. Strominger, 
H. Verlinde, E. Witten and Z. Yin for helpful discussions and comments.
KD would also like to thank the string theory group
at Harvard University, where part of this work was done, for stimulating
discussions.
The research of OJG is supported by NSF grant no.~PHY-9802484;
the research of JLK is in part supported by the Natural Sciences and
Engineering Research Council of Canada; the research of KD is supported
by Department of Energy grant no.~DE-FG02-90ER40542
and the research of GR is supported by NSF grant no.~PHY-0070928 and by
a Helen and Martin Chooljian fellowship.

\appendix
\section{Generic Orientation of the Dipole Vectors}\label{appthree}

In section (\ref{sugsol}) we promised to describe the supergravity
solution for generic dipole theories where the various dipole vectors
are not all along the same direction.  In order to avoid clutter, we will
set $\a' = 1$ in this appendix.

We start with a D3-brane extended in the 0123 directions,
compactified on a $T^3$ with radii $(R_1, R_2, R_3).$
The relevant non-zero fields are
\bear
ds^2 &=& \frac{1}{\sqrt{H}}
\left (
dt^2 - (R_1 dx^1)^2 - (R_2 dx^2)^2 - (R_3 dx^3)^2
\right )
- \sqrt{H} (dx^a)^2
\nn\\
C_{0123} &=& - \frac{1}{H}
\nn\\
\varphi &=& \varphi_0
\label{D3}
\eear
where
$$
H = 1 + \frac {{\mathcal{R}}^4}{r^4} \qquad \qquad
r^2 \equiv (x^a)^2
$$
The roman indices $a,b, \dots$ run from 4 to 9, and
we use greek indices to indicate the compactified
directions 1,2,3.
Starting from the solution (\ref{D3}), we perform
the T-duality transformation three times, in the three
compactified directions using the formulae
of \cite{BUSH,BUSC,BHO}.  The answer, which is a D0-brane smeared
over the T-dual torus $T^3: (R_1^{-1},R_2^{-1},R_3^{-1})$, is
\bear
ds^2 &=& \frac{1}{\sqrt{H}} dt^2
- \sqrt{H} \left (
\left (\frac{dx^1}{R_1}\right)^2 +
\left (\frac{dx^2}{R_2}\right)^2 +
\left (\frac{dx^3}{R_3}\right)^2
\right )
- \sqrt{H} (dx^a)^2
\nn\\
C^{(1)}_{0} &=& - \frac{4}{H}
\nn\\
e^{2(\phi - \varphi_0)} &=& \frac {H^{3/2}}{R_1^2 R_2^2 R_3^2}
\label{D0}
\eear
Now, we introduce the three twists, by replacing
$$
dx^a \longrightarrow dx^a - \sum_\mu
(\Omega^{\mu}_{ab} x^b) dx^{\mu}
$$
where $(\Omega^{\mu})^\top = -\Omega^{\mu}$ are commuting elements of SO(6).
The metric with the twist is
\be
ds^2 = \frac{1}{\sqrt{H}} dt^2
- \sqrt{H} \left (
\left (\frac{dx^1}{R_1}\right)^2 +
\left (\frac{dx^2}{R_2}\right)^2 +
\left (\frac{dx^3}{R_3}\right)^2
\right )
- \sqrt{H} \left( dx^a - (\Omega^{\mu}_{ab} x^b) dx^{\mu}\right)^2
\label{gentwmet}
\label{D0twist}
\ee
Now, we T-dualize three times
to get back the metric for a D3-brane with a dipole theory living on it.
Define $M^{\mu} \equiv R_{\mu} \Omega^{\mu}$
(no contraction over $\mu$) and $x^a \equiv r \nv^a$
where $\nv^\top \nv = 1$. With some work, the metric turns out to be
(here and below there is no contraction in terms like
$R_\v dx^\v$)
\bear
ds^2 &=& \frac {1}{\sqrt{H}} dt^2 - \sqrt{H} dr^2
\nn\\
&-&\frac {1}{\sqrt{H}}
\frac
{
\epsilon_{\alpha\beta\gamma} \epsilon_{\kappa\mu\nu}
\left [\delta^{\alpha\kappa} + r^2 (m^\alpha)^\top m^\kappa \right]
\left [\delta^{\beta\mu} + r^2 (m^\beta)^\top m^\mu \right]
}
{2D}
(R_\gamma dx^\gamma)(R_\nu dx^\nu)
\nn\\
&-& \sqrt{H} (r^2 dn^T dn)
\nn\\
&+& \sqrt{H} \left (
\frac
{r^4
\epsilon_{\alpha\beta\gamma} \epsilon_{\kappa\mu\nu}
\left [\delta^{\alpha\kappa} + r^2 (m^\alpha)^\top m^\kappa \right]
\left [\delta^{\beta\mu} + r^2 (m^\beta)^\top m^\mu \right]
\left [ ((m^\gamma)^\top dn) ((m^\nu)^\top dn)  \right ]
}
{2D}
\right ) \nn
\eear
where we have defined
$$
D \equiv \frac{1}{6}
\epsilon_{\alpha\beta\gamma} \epsilon_{\kappa\mu\nu}
\left [\delta^{\alpha\kappa} + r^2 (m^\alpha)^\top m^\kappa \right]
\left [\delta^{\beta\mu} + r^2 (m^\beta)^\top m^\mu \right]
\left [\delta^{\gamma\nu} + r^2 (m^\gamma)^\top m^\nu \right]
$$
and
$$
   m^\alpha \equiv M^\alpha \nv
$$
The other nonzero fields are
\bear
C^{(4)}_{0123} &=& \frac{1}{H}
\nn\\
B^{(1)}_{\mu a} dx^\mu \wdg d\nv^a &=& - \sqrt{H} j_{\gamma\nu}
dx^\gamma \wdg \frac {r m^{\mu} d\nv}{R_\nu}
\\ &=&
  \frac{r
\epsilon_{\alpha\beta\gamma} \epsilon_{\kappa\mu\nu}
\left [\delta^{\alpha\kappa} + r^2 (m^\alpha)^\top m^\kappa \right]
\left [\delta^{\beta\mu} + r^2 (m^\beta)^\top m^\mu \right]
\left [(R_\gamma dx^\gamma) \wdg ((m^\nu)^\top d\nv)  \right ]
}{2 D}
\nn\\
e^{2(\varphi - \varphi_0)} &=& \frac{1}{D}
\nn
\label{D3twist}
\eear

For $M^1=M^2=0$ this reduces to the answers for
a single twist. It is interesting to ask what happens when the
twists, $M^{\u}$, do not commute. In this situation, the Ricci scalar
of the twisted metric (\ref{gentwmet}) has a field strength term, and
thus the metric is no longer a solution to the supergravity equations.

\def\np#1#2#3{{\it Nucl.\ Phys.} {\bf B#1} (#2) #3}
\def\pl#1#2#3{{\it Phys.\ Lett.} {\bf B#1} (#2) #3}
\def\physrev#1#2#3{{\it Phys.\ Rev.\ Lett.} {\bf #1} (#2) #3}
\def\prd#1#2#3{{\it Phys.\ Rev.} {\bf D#1} (#2) #3}
\def\ap#1#2#3{{\it Ann.\ Phys.} {\bf #1} (#2) #3}
\def\ppt#1#2#3{{\it Phys.\ Rep.} {\bf #1} (#2) #3}
\def\rmp#1#2#3{{\it Rev.\ Mod.\ Phys.} {\bf #1} (#2) #3}
\def\cmp#1#2#3{{\it Comm.\ Math.\ Phys.} {\bf #1} (#2) #3}
\def\mpla#1#2#3{{\it Mod.\ Phys.\ Lett.} {\bf #1} (#2) #3}
\def\jhep#1#2#3{{\it JHEP.} {\bf #1} (#2) #3}
\def\atmp#1#2#3{{\it Adv.\ Theor.\ Math.\ Phys.} {\bf #1} (#2) #3}
\def\jgp#1#2#3{{\it J.\ Geom.\ Phys.} {\bf #1} (#2) #3}
\def\cqg#1#2#3{{\it Class.\ Quant.\ Grav.} {\bf #1} (#2) #3}
\def\hepth#1{{\tt hep-th/{#1}}}


\end{document}